\documentstyle[12pt]{article}
\newlength{\extraspace}
\newlength{\extraspaces}
\newcommand{\be}{\begin{equation}
\addtolength{\abovedisplayskip}{\extraspaces}
\addtolength{\belowdisplayskip}{\extraspaces}
\addtolength{\abovedisplayshortskip}{\extraspace}
\addtolength{\belowdisplayshortskip}{\extraspace}}
\newcommand{\ee}{\end{equation}}

\newcommand{\ba}{\begin{eqnarray}
\addtolength{\abovedisplayskip}{\extraspaces}
\addtolength{\belowdisplayskip}{\extraspaces}
\addtolength{\abovedisplayshortskip}{\extraspace}
\addtolength{\belowdisplayshortskip}{\extraspace}}
\newcommand{\ea}{\end{eqnarray}}

\newcommand{\newsection}[1]{
\vspace{15mm}
\pagebreak[3]
\addtocounter{section}{1}
\setcounter{equation}{0}
\setcounter{subsection}{0}
\setcounter{footnote}{0}
\begin{flushleft}
{\large\bf \thesection. #1}
\end{flushleft}
\nopagebreak
\medskip
\nopagebreak}

\newcommand{\Tr}{{\rm Tr}}

\begin{document}

\addtolength{\baselineskip}{.8mm}

{\thispagestyle{empty}
\noindent \hspace{1cm}  \hfill February 1996 \hspace{1cm}\\
\mbox{}                 \hfill IFUP--TH 10/96 \hspace{1cm}\\

\begin{center}\vspace*{1.0cm}
{\large\bf The high--energy quark--quark scattering:}\\
{\large\bf from Minkowskian to Euclidean theory}\\
\vspace*{1.0cm}
{\large Enrico Meggiolaro}\\
\vspace*{0.5cm}{\normalsize
{Dipartimento di Fisica, \\
Universit\`a di Pisa, \\ 
I--56100 Pisa, Italy.}}\\
\vspace*{2cm}{\large \bf Abstract}
\end{center}

\noindent
In this paper we consider some analytic properties of the high--energy 
quark--quark scattering amplitude, which, as is well known, can be
described by the expectation value of two lightlike Wilson lines, running 
along the classical trajectories of the two colliding particles.
We shall prove that the expectation value of two infinite Wilson lines,
forming a certain hyperbolic angle in Minkowski space--time, and the 
expectation value of two infinite Euclidean Wilson lines, forming a certain 
angle in Euclidean four--space, are connected by an analytic
continuation in the angular variables. This could open the possibility of
evaluating the high--energy scattering amplitude directly on the lattice or 
using the stochastic vacuum model.
The Abelian case (QED) is also discussed.
}
\vfill\eject

\newsection{Introduction}

\noindent
There is a class of {\it soft} high--energy scattering processes,
i.e., elastic scattering processes at high squared energies $s$ in 
the center of mass and small squared transferred momentum $t$ (that is
$s \to \infty$ and $|t| \ll s$, let us say $|t| \le 1~{\rm GeV}^2$), for which
QCD perturbation theory cannot be safely applied, since $t$ is too small.
Elaborate procedures for summing perturbative contributions have been 
developed \cite{Cheng-Wu-book} \cite{Lipatov}, even if the results are not 
able to explain the most relevant phenomena.

A non--perturbative analysis, based on QCD, of these high--energy 
scattering processes was performed by Nachtmann in \cite{Nachtmann91}.
He studied the $s$ dependence of the quark--quark (and quark--antiquark) 
scattering amplitude by analytical means, using a functional integral 
approach and an eikonal approximation to the solution of the Dirac equation 
in the presence of a non--Abelian external gluon field.

In a previous paper \cite{Meggiolaro95} we proposed an approach to 
high--energy quark--quark (and quark--antiquark) scattering, 
based on a first--quantized path--integral 
description of quantum field theory developed by Fradkin in the early 1960's
\cite{Fradkin}. In this approach one obtains convenient expressions for the 
full and truncated--connected scalar propagators in an external 
(gravitational, electromagnetic, etc.) field and the eikonal approximation 
can be easily recovered in the relevant limit. Knowing the 
truncated--connected propagators, one can then extract, in the manner of 
Lehmann, Symanzik, and Zimmermann (LSZ),
the scattering matrix elements in the framework of a functional integral 
approach. This method was originally adopted
in \cite{Veneziano} in order to study Planckian--energy 
gravitational scattering. 

The high--energy quark--quark scattering amplitude comes out to be
described by the expectation value of two lightlike Wilson lines, running 
along the classical trajectories of the two colliding particles.

In the center--of--mass reference system (c.m.s.), taking the initial 
trajectories of the two quarks along the $x^1$--axis, the initial four--momenta
$p_1$, $p_2$ and the final four--momenta $p'_1$ and $p'_2$ are given, 
in the first approximation ({\it eikonal} approximation), by
\be
p_1 \simeq p'_1 \simeq (E,E,{\bf 0}_t) ~~~ , ~~~ 
p_2 \simeq p'_2 \simeq (E,-E,{\bf 0}_t) ~.
\ee
Let us indicate with $x_1^\mu (\tau)$ and $x_2^\mu (\tau)$ the classical 
trajectories of the two colliding particles in Minkowski space--time:
\be
x_1^\mu (\tau) = z_t^\mu + p_1^\mu \tau ~~~ , ~~~
x_2^\mu (\tau) = p_2^\mu \tau ~,
\ee
where $z_t = (0,0,{\bf z}_t)$, with ${\bf z}_t = (z^2,z^3)$, is the 
distance between the two trajectories in the {\it transverse} plane
(the coordinates $(x^0,x^1)$ are often called {\it longitudinal} coordinates).
The high--energy ($s \to \infty$ and $|t| \ll s$) quark--quark scattering 
amplitude turns out to be controlled by the Fourier transform,
with respect to the transverse coordinates ${\bf z}_t$,
of the expectation value of the two lightlike Wilson lines running along
$x_1^\mu (\tau)$ and $x_2^\mu (\tau)$:
\ba
W_1 (z_t) &=&
{P} \exp \left[ -ig \displaystyle\int_{-\infty}^{+\infty}
A_\mu (z_t + p_1 \tau) p_1^\mu d\tau \right] ~;
\nonumber \\
W_2 (0) &=&
{P} \exp \left[ -ig \displaystyle\int_{-\infty}^{+\infty}
A_\mu (p_2 \tau) p_2^\mu d\tau \right] ~,
\ea
where $P$ stands for ``{\it path ordering}'' and $A_\mu = A_\mu^a T^a$.
The space--time configuration of these two Wilson lines is shown in Fig. 1.

\newcommand{\wilson}[1]{
\begin{figure}
\begin{center}
\setlength{\unitlength}{1.00mm}
\raisebox{-40\unitlength}
{\mbox{\begin{picture}(80,45)(-35,-30)
\thicklines
\put(-22,22){\line(1,-1){41}}
\put(-15,15){\vector(-1,1){1}}
\put(16,23){\line(-1,-1){42}}
\put(9,16){\vector(1,1){1}}
\put(0,0){\vector(-2,-1){13}}
\thinlines
\put(-8,0){\line(1,0){35}}
\put(8,4){\line(-2,-1){35}}
\put(0,-8){\line(0,1){35}}
\put(27,0){\vector(1,0){1}}
\put(0,27){\vector(0,1){1}}
\put(-13,20){\makebox(0,0){$W_2$}}
\put(18,20){\makebox(0,0){$W_1$}}
\put(-6,-7){\makebox(0,0){$z_t$}}
\put(25,-2){\makebox(0,0){$x$}}
\put(-2,25){\makebox(0,0){$t$}}
\end{picture}}}
\parbox{13cm}{\small #1}
\end{center}
\end{figure}}

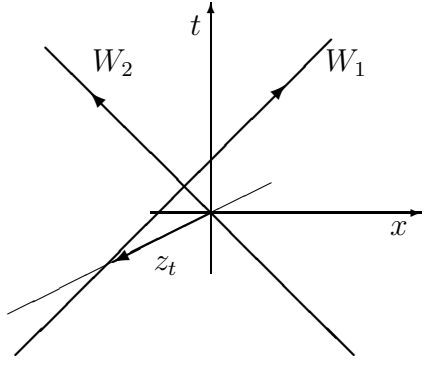
\wilson{{\bf Fig.~1.} The space--time configuration of the two lightlike
Wilson lines $W_1$ and $W_2$ entering in the expression (1.4) for the
high--energy quark--quark elastic scattering amplitude.}

Explicitly indicating the color indices ($i,j, \ldots$)
and the spin indices ($\alpha, \beta, \ldots$) of the quarks, 
the scattering amplitude can be written as
\ba
\lefteqn{
M_{fi} = \langle \psi_{i\alpha}(p'_1) \psi_{k\gamma}(p'_2) | M | 
\psi_{j\beta}(p_1) \psi_{l\delta}(p_2) \rangle } \nonumber \\
& & \mathop{\sim}_{s \to \infty}
-{i \over Z_\psi^2} \cdot \delta_{\alpha\beta} \delta_{\gamma\delta}
\cdot 2s 
\displaystyle\int d^2 {\bf z}_t e^{i {\bf q} \cdot {\bf z}_t}
\langle [ W_1 (z_t) - {\bf 1} ]_{ij} [ W_2 (0) - {\bf 1} ]_{kl} \rangle_A 
~,
\ea
where $\langle \ldots \rangle_A$ is the average, in the sense of the 
functional integration, over the gluon field $A^\mu$.
$Z_\psi$ is the fermion--field renormalization constant, which can be 
written in the eikonal approximation as \cite{Nachtmann91}
\be
Z_\psi \simeq {1 \over N_c} \langle \Tr [ W_1 (z_t) ] \rangle_A 
= {1 \over N_c} \langle \Tr [ W_1 (0) ] \rangle_A 
= {1 \over N_c} \langle \Tr [ W_2 (0) ] \rangle_A ~.
\ee
(The two last equalities come from the Poincar\'e invariance of the theory.)

In a perfectly analogous way one can also derive the high--energy 
scattering amplitude in the case of the Abelian group $U(1)$ (QED).
The resulting amplitude is equal to Eq. (1.4), with the only obvious 
difference being that now the lightlike Wilson lines $W_1$ and $W_2$ are 
functionals of the Abelian field $A^\mu$ (so they are not matrices).
Thanks to the simple form of the Abelian theory (in particular to the 
absence of self--interactions among the vector fields), it turns out that 
it is possible to explicitly evaluate (at least in the {\it quenched} 
approximation) the expectation value of the two 
Wilson lines: the details of the calculation are reported in the Appendix 
of Ref. \cite{Meggiolaro95}
and one finally recovers the well--known result for the eikonal amplitude 
of the high--energy scattering in QED \cite{Cheng-Wu} 
\cite{Abarbanel-Itzykson} \cite{Jackiw}.

From Eq. (1.4) it seems that the 
$s$ dependence of the scattering amplitude is all contained in the 
kinematic factor $2s$ in front of the integral. In fact we can write
\be
M_{fi} = \langle \psi_{i\alpha}(p'_1) \psi_{k\gamma}(p'_2) | M | 
\psi_{j\beta}(p_1) \psi_{l\delta}(p_2) \rangle
\mathop{\sim}_{s \to \infty}
-i \cdot 2s \cdot \delta_{\alpha\beta} \delta_{\gamma\delta}
\cdot g_{M (ij,kl)} (t,s) ~,
\ee
where, apparently, the quantity
\be
g_{M (ij,kl)} (t,s) \equiv {1 \over Z_\psi^2}
\displaystyle\int d^2 {\bf z}_t e^{i {\bf q} \cdot {\bf z}_t}
\langle [ W_1 (z_t) - {\bf 1} ]_{ij} [ W_2 (0) - {\bf 1} ]_{kl} \rangle_A ~
\ee
only depends on $t = -{\bf q}^2$. Yet, as was pointed 
out by Verlinde and Verlinde in \cite{Verlinde}, this is not true:
in fact, one can easily be convinced (for example by making a perturbative 
expansion) that it is a singular limit to take the Wilson lines in
(1.7) exactly lightlike. As suggested in \cite{Verlinde}, one can 
regularize this sort of ``infrared'' divergence by letting each line
have a small timelike component, so that they coincide with the classical 
trajectories for quarks with a finite mass $m$. Therefore, one first has to 
evaluate the quantity
\be
g_{M (ij,kl)} (t, \beta)
\ee
for two Wilson lines along the trajectories of two quarks moving with 
velocity $\beta$ and $-\beta$ ($0 <  \beta <  1$) along the $x^1$--axis.
In other words, one first considers two infinite Wilson lines forming a 
certain hyperbolic angle $\chi$ in Minkowski space--time.
Then, to obtain the correct high--energy scattering amplitude, one has to 
perform the limit $\beta \to 1$, that is $\chi \to \infty$, into the 
expression (1.8):
\be
M_{fi} = \langle \psi_{i\alpha}(p'_1) \psi_{k\gamma}(p'_2) | M | 
\psi_{j\beta}(p_1) \psi_{l\delta}(p_2) \rangle
\mathop{\sim}_{s \to \infty}
-i \cdot 2s \cdot \delta_{\alpha\beta} \delta_{\gamma\delta}
\cdot g_{M (ij,kl)} (t, \beta \to 1) ~.
\ee
In this way one obtains 
a $\ln s$ dependence of the amplitude, as expected from ordinary 
perturbation theory \cite{Cheng-Wu-book} \cite{Lipatov} and as confirmed by 
the experiments on hadron--hadron scattering processes.
In Sect. 3 we shall see how this explicitly works by evaluating
the amplitude (1.7) for QCD up to the order $O(g_R^4)$ in the perturbative 
expansion ($g_R$ being the renormalized coupling constant).

The direct evaluation of the expectation value (1.7) is a 
highly non--trivial matter, as it is also strictly connected with the
ultraviolet properties of Wilson--line operators \cite{Arefeva80}.
Recently, in Ref. \cite{Korchemsky}, it has been found that there is a 
correspondence between high--energy asymptotics in QCD and renormalization 
properties of the so--called cross singularities of Wilson lines.
The asymptotic behavior of the quark--quark scattering amplitude turns out 
to be controlled by a $2 \times 2$ matrix of the cross anomalous dimensions 
of Wilson lines. 
An alternative non--perturbative approach for the calculation of the 
expectation value (1.7) has been proposed in Ref. \cite{Arefeva94}.
It consists in studying the Regge regime of large energies and fixed 
momentum transfers as a special regime of lattice gauge theory on an 
asymmetric lattice, with a spacing $a_0$ in the longitudinal direction and 
a spacing $a_t$ in the transverse direction, in the limit $a_0 / a_t
\to 0$.

At the moment, the only non--perturbative numerical estimate of (1.7), 
which can be found in the literature, is that of Ref. \cite{Dosch} (where
it has been generalized to the case of hadron--hadron scattering): it has 
been obtained in the framework of the model of the stochastic vacuum (SVM).
Before its application to high--energy scattering, the SVM must be 
translated from Euclidean space--time, in which it is naturally formulated, 
to the Minkowski continuum. As is claimed in Ref. \cite{Dosch}, the more 
safe way (from the point of view of the functional integration) would be 
the other way, i.e., to continue the scattering amplitude from the 
Minkowski world to the Euclidean world. 

In this paper we try to go just that way and adapt the scattering amplitude 
to the Euclidean world. More explicitly, we shall prove that the 
expectation value of two infinite Wilson lines,
forming a certain hyperbolic angle in Minkowski space--time, and the 
expectation value of two infinite Euclidean Wilson lines, forming a certain 
angle in Euclidean four--space, are connected by an analytic
continuation in the angular variables. In Sect. 2 we shall first prove this 
for the Abelian case (QED), by explicitly evaluating the correlation of two 
infinite Wilson lines both in Minkowski space--time and in
Euclidean four--space, using the so--called {\it quenched} approximation.
Then, in Sect. 3, we shall prove that this result can be extended also to 
non--Abelian gauge theories.
Finally, in the last section, we shall discuss some interesting 
consequences (such as the re--derivation of the {\it Regge pole model}
\cite{Regge})
and some possible direct applications, mostly for lattice gauge theories (LGT)
and the stochastic vacuum model, of this relationship of analytic
continuation.

\newsection{The Abelian case}

\noindent
In this section we shall discuss the Abelian case (See also 
Refs. \cite{Korchemsky} and \cite{Arefeva94}). 
The fermion--fermion electromagnetic scattering 
amplitude, in the high--energy limit $s \to \infty$ and $|t| \ll s$, can be 
derived following the same procedure used in Ref. \cite{Meggiolaro95}.
The resulting amplitude 
is formally identical to Eq. (1.4), with the only obvious 
difference being that now the Wilson lightlike lines $W_1$ and $W_2$ are 
functions of the Abelian field $A^\mu$ (so they are not matrices):
\ba
\lefteqn{
M_{fi} = \langle \psi_{\alpha}(p'_1) \psi_{\gamma}(p'_2) | M | 
\psi_{\beta}(p_1) \psi_{\delta}(p_2) \rangle } \nonumber \\
& & \mathop{\sim}_{s \to \infty}
-{i \over Z_\psi^2} \cdot \delta_{\alpha\beta} \delta_{\gamma\delta}
\cdot 2s
\displaystyle\int d^2 {\bf z}_t e^{i {\bf q} \cdot {\bf z}_t}
\langle [ W_1 (z_t) - 1 ] [ W_2 (0) - 1 ] \rangle_A ~.
\ea
The electromagnetic lightlike Wilson lines $W_1$ and $W_2$ are defined 
as in (1.3),
after replacing $g$ with $e$, the electric coupling--constant (electric 
charge), and the gluon field with the photon field.
Thanks to the simple form of the Abelian theory (in particular to the 
absence of self--interactions among the vector fields), it turns out that 
it is possible to explicitly evaluate the expectation value of the two 
Wilson lines, thus finally recovering the well--known result for the eikonal 
amplitude of the high--energy scattering in QED (see Refs. \cite{Cheng-Wu},
\cite{Abarbanel-Itzykson} and \cite{Jackiw}). 
The details of the calculation are reported in Ref. \cite{Meggiolaro95}.

The Wilson lines in (2.1) are taken exactly lightlike. 
We shall now let each line to
have a small timelike component, so that they coincide with the classical 
trajectories for fermions with a finite mass $m$. The electromagnetic lightlike 
Wilson lines $W_1$ and $W_2$ are now defined as
\ba
W_1 (z_t) &=&
\exp \left[ -ie \displaystyle\int_{-\infty}^{+\infty}
A_\mu (z_t + {p_1 \over m} \tau) {p_1^\mu \over m} d\tau \right] ~,
\nonumber \\
W_2 (0) &=&
\exp \left[ -ie \displaystyle\int_{-\infty}^{+\infty}
A_\mu ({p_2 \over m} \tau) {p_2^\mu \over m} d\tau \right] ~,
\ea
where $m$ is the mass of the fermions and $p_1^\mu$ and $p_2^\mu$ are the 
two four--momenta defining the two trajectories 1 and 2 in Minkowski 
space--time:
\ba
X^\mu_{(1)} (\tau) &=& z_t^\mu + {p_1^\mu \over m} \tau ~, \nonumber \\
X^\mu_{(2)} (\tau) &=& {p_2^\mu \over m} \tau ~.
\ea
In the c.m.s. of the two particles, taking the 
spatial momenta ${\bf p}_1$ and ${\bf p}_2 = -{\bf p}_1$ along the 
$x^1$--direction, the two four--momenta $p_1$ and $p_2$ are
\ba
p_1^\mu &=& E (1,\beta,{\bf 0}_t) ~, \nonumber \\
p_2^\mu &=& E (1,-\beta,{\bf 0}_t) ~, 
\ea
where $\beta$ is the velocity (in the units with $c=1$) and 
$E = m / \sqrt{1 - \beta^2}$ is the 
energy of each particle (so that: $s = 4E^2$).

We shall evaluate the expectation value 
$\langle W_1 (z_t) W_2 (0) \rangle_A$ in the
so--called {\it quenched} approximation, where vacuum polarization effects, 
arising from the presence of loops of dynamical fermions, are neglected.
This amounts to setting $\det (K[A]) = 1$, where $K[A] = i\gamma^\mu D_\mu
- m$ is the fermion matrix.
Thus we can write that
\be
\langle W_1 (z_t) W_2 (0) \rangle_A \simeq
{1 \over Z} \displaystyle\int [dA_\mu] e^{iS_A} W_1 (z_t) W_2 (0) ~,
\ee
where $S_A = -{1 \over 4} \displaystyle\int d^4 x F_{\mu\nu} F^{\mu\nu}$ 
is the action 
of the electromagnetic field and $Z = \displaystyle\int [dA_\mu] e^{iS_A}$ 
is the pure--gauge partition function.
We then add to the pure--gauge Lagrangian $L_A = -{1 \over 4} F_{\mu\nu}
F^{\mu\nu}$ a gauge--fixing term $L_{GF} = -{1 \over 2\alpha}
(\partial^\mu A_\mu)^2$ ({\it covariant} or {\it Lorentz} gauge). The 
expectation value (2.5) becomes, denoting $L_0^F = L_A + L_{GF}$,
\be
\langle W_1 (z_t) W_2 (0) \rangle_A \simeq
{1 \over Z'} \displaystyle\int [dA]
\exp \left[ i \displaystyle\int d^4 x ( L_0^F + A_\mu J^\mu ) \right] ~,
\ee
where $Z' = \displaystyle\int [dA] \exp \left( i\displaystyle\int d^4 x L_0^F
\right)$ and $J^\mu (x)$ is a four--vector source defined as
\be
J^\mu (x) = J^\mu_{(1)} (x) + J^\mu_{(2)} (x) \nonumber \\
= -e [ \epsilon^\mu_{(1)} \delta (x - \beta t) \delta ({\bf x}_t - {\bf z}_t)
+ \epsilon^\mu_{(2)} \delta (x + \beta t) \delta ({\bf x}_t) ] ~,
\ee
with $\epsilon^\mu_{(1)} = (1,\beta,0,0)$ and $\epsilon^\mu_{(2)} =
(1,-\beta,0,0)$.
The functional integral
\be
Z_0 [J] \equiv \displaystyle\int [dA] \exp 
\left[ i\displaystyle\int d^4 x ( L_0^F + A_\mu J^\mu ) \right]
\ee
can be evaluated with standard methods (completing the quadratic form in 
the exponent). One thus obtains
\be
Z_0 [J] = Z' \cdot \exp \left[ {i \over 2} \displaystyle\int d^4 x 
\displaystyle\int d^4 y J^\mu (x) D_{\mu\nu} (x-y) J^\nu (y) \right] ~,
\ee
where $D_{\mu\nu}$ is the free photon propagator (apart from a factor $-i$):
\be
D_{\mu\nu} (x-y) = \displaystyle\int {d^4 k \over (2\pi)^4}
{e^{-ik(x-y)} \over k^2 +i\varepsilon}
\left( g_{\mu\nu} - (1 - \alpha){k_\mu k_\nu \over k^2
+ i\varepsilon} \right) ~.
\ee
In the following we shall choose the gauge--fixing parameter $\alpha$ equal 
to $1$ ({\it Feynman} gauge). 
From Eqs. (2.6) $\div$ (2.10), we derive the following expression 
for the expectation value (2.5) of the two Wilson lines:
\ba
\lefteqn{
\langle W_1 (z_t) W_2 (0) \rangle_A \simeq 
\langle W_1 (z_t) \rangle_A \langle W_2 (0) \rangle_A }
\nonumber \\
& & \times \exp \left[ i \displaystyle\int d^4 x \displaystyle\int d^4 y 
J_{(1)}^\mu (x) J_{(2)\mu} (y) \displaystyle\int {d^4 k \over (2\pi)^4}
{e^{-ik(x-y)} \over k^2 +i\varepsilon} \right] ~.
\ea
Therefore, using the explicit form (2.7) of the four--vector source $J^\mu (x)$ 
to evaluate the double integral in Eq. (2.11), one finds that (in the 
{\it quenched} approximation)
\be
{ \langle W_1 (z_t) W_2 (0) \rangle_A \over 
\langle W_1 \rangle_A^2 }
\simeq \exp \left[ -i e^2 \left( { 1 + \beta^2 \over 2\beta } \right)
\displaystyle\int {d^2 {\bf k}_t \over (2\pi)^2}
{e^{-i{\bf k}_t \cdot {\bf z}_t} \over {\bf k}_t^2 - i\varepsilon} \right] 
~.
\ee
We have made use of the Poincar\'e invariance of the theory to write:
$\langle W_1 \rangle_A \equiv \langle W_1 (z_t) \rangle_A =
\langle W_1 (0) \rangle_A = \langle W_2 (0) \rangle_A$.
We now introduce the hyperbolic angle $\psi$ [in the plane $(x^0,x^1)$]
of the trajectory 1 in Eq. (2.3), i.e., $\Delta X_{(1)}^1 = \beta 
\Delta X_{(1)}^0$, so that, if $\Delta l$ is the line--distance 
($\Delta l^2 = (\Delta X_{(1)}^0)^2 -(\Delta X_{(1)}^1)^2 
= (\Delta X_{(1)}^0)^2 (1 - \beta^2)$) we have
\ba
\Delta X_{(1)}^0 &=& \Delta l \cosh \psi ~, \nonumber \\
\Delta X_{(1)}^1 &=& \Delta l \sinh \psi ~,
\ea
and, therefore, $\beta = \tanh \psi$. With this notation, it is immediate 
to recognize that the $\beta$--dependent factor in front of Eq. (2.12) is 
equal to $\coth (2\psi)$; so that
\be
{ \langle W_1 (z_t) W_2 (0) \rangle_A \over \langle W_1 \rangle_A^2 }
\simeq \exp \left[ -i e^2 \coth \chi \cdot
\displaystyle\int {d^2 {\bf k}_t \over (2\pi)^2}
{e^{-i{\bf k}_t \cdot {\bf z}_t} \over {\bf k}_t^2 - i\varepsilon} \right] 
~,
\ee
where $\chi = 2\psi$ is hyperbolic angle [in the plane $(x^0,x^1)$] 
between the two trajectories 1 and 2 of the two colliding particles.
The exponential in the last equation turns out to be equal to
\be
e^{-ie^2 \Lambda \coth \chi} \cdot 
\exp \left( i {e^2 \over 2\pi} \coth \chi \cdot \ln |{\bf z}| \right) ~,
\ee
where $\Lambda$ is an infinite constant phase and is therefore physically 
unobservable. The origin of this infinite constant phase resides in the 
fact that the fermion--fermion scattering amplitude in QED has 
infrared (IR) divergences, due to the emission of low--energy massless vector 
mesons. The traditional way to handle these IR divergences is to 
introduce an IR {\it cutoff} in the form of a vector meson mass $\lambda$.
In this way the integral over ${\bf k}_t$ in the exponent of Eq. (2.14) is 
substituted by the expression
\be
\displaystyle\int {d^2 {\bf k}_t \over (2\pi)^2}
{e^{-i{\bf k}_t \cdot {\bf z}_t} \over {\bf k}_t^2 + \lambda^2} \equiv
{1 \over 2\pi} K_0 (\lambda |{\bf z}_t|) ~,
\ee
where $K_0$ is the modified Bessel function. In the limit of small $\lambda$
this last expression can be replaced by
\be
{1 \over 2\pi} K_0 (\lambda |{\bf z}|) \mathop{\sim}_{\lambda \to 0}
-{1 \over 2\pi} \ln \left( {1 \over 2} e^\gamma \lambda |{\bf z}_t| \right) ~.
\ee
Absorbing ${1 \over 2} e^\gamma$ in $\lambda$ and putting 
$\Lambda = -(1/2\pi) \ln ({1 \over 2} e^\gamma \lambda)$ (so that $\Lambda
\to \infty$ when $\lambda \to 0$), we just obtain the expression (2.15) for
the exponential in Eq. (2.14).

We can now repeat the above procedure and evaluate the 
quantity (2.5) in the Euclidean theory.
The electromagnetic lightlike Euclidean Wilson lines $W_{E 1}$ and 
$W_{E 2}$ are defined as in (2.2):
\ba
W_{E 1} (\bar{z}_{t}) &=&
\exp \left[ -ie \displaystyle\int_{-\infty}^{+\infty}
A^{(E)}_{ \mu} (\bar{z}_{t} + v_1 \tau) v_{1\mu} d\tau \right] ~,
\nonumber \\
W_{E 2} (0) &=&
\exp \left[ -ie \displaystyle\int_{-\infty}^{+\infty}
A^{(E)}_{ \mu} (v_2 \tau) v_{2\mu} d\tau \right] ~,
\ea
where now $v_{1 \mu}$ and $v_{2 \mu}$ are the Euclidean four--vectors 
[lying in the plane $(x_1,x_4)$]
defining the two trajectories 1 and 2 in Euclidean four--space:
\ba
X^{(1)}_{E \mu} (\tau) &=& \bar{z}_{t \mu} + v_{1 \mu} \tau ~, \nonumber \\
X^{(2)}_{E \mu} (\tau) &=& v_{2 \mu} \tau ~,
\ea
and $\bar{z}_{t \mu} = (z_1, z_2, z_3, z_4) = (0, {\bf z}_t, 0)$.
We can choose $v_1$ and $v_2$ normalized to 1: $v_1^2 = v_2^2 = 1$.
Moreover, due to the $O(4)$ symmetry of the theory, we can choose
the c.m.s. of the two particles, taking the 
spatial momenta ${\bf v}_1$ and ${\bf v}_2 = -{\bf v}_1$ along the 
$x_1$--direction. The two four--momenta $v_1$ and $v_2$ are, therefore,
\ba
v_{1 \mu} &=& (\sin \phi,{\bf 0}_t,\cos \phi) ~, \nonumber \\
v_{2 \mu} &=& (-\sin \phi,{\bf 0}_t,\cos \phi) ~, 
\ea
where $\phi$ is the angle formed by each trajectory with the $x_4$--axis.
As before, we can evaluate the expectation value 
$\langle W_{E 1} (\bar{z}_{t}) W_{E 2} (0) \rangle_A$ (where now 
$\langle \ldots \rangle_A$ is the 
functional integral with respect to the gluon field $A^{(E)}_{ \mu}$ in the 
Euclidean theory) in the {\it quenched} approximation.
Thus we have
\be
\langle W_{E 1} (\bar{z}_{t}) W_{E 2} (0) \rangle_A \simeq
{1 \over Z_E} \displaystyle\int [dA^{(E)}_{ \mu}] 
e^{-S^{(E)}_{ A}} W_{E 1} (\bar{z}_{t}) W_{E 2} (0) ~,
\ee
where $S^{(E)}_{A} = {1 \over 4} \displaystyle\int d^4 x_E F^{(E)}_{ \mu\nu} 
F^{(E)}_{ \mu\nu}$ is the Euclidean action of the electromagnetic field and 
$Z_E = \displaystyle\int [dA^{(E)}_{ \mu}] e^{-S^{(E)}_{A}}$ is the 
corresponding pure--gauge partition function.
As usually, we add to the pure--gauge Lagrangian $L^{(E)}_{A} = {1 \over 4} 
F^{(E)}_{ \mu\nu} F^{(E)}_{ \mu\nu}$ a gauge--fixing term 
$L^{(E)}_{GF} = {1 \over 2\alpha} (\partial_\mu A^{(E)}_{ \mu})^2$ 
({\it covariant} or {\it Lorentz} gauge). The expectation value (2.21) 
becomes, denoting $L^{(E)}_{0} = L^{(E)}_{A} + L^{(E)}_{GF}$,
\be
\langle W_{E 1} (\bar{z}_{t}) W_{E 2} (0) \rangle_A \simeq
{1 \over Z'_E} \displaystyle\int [dA^{(E)}_{}] 
\exp \left[ - \displaystyle\int d^4 x_E 
( L^{(E)}_{0} + i A^{(E)}_{ \mu} J_{E  \mu} ) \right] ~,
\ee
where $Z'_E = \displaystyle\int [dA^{(E)}_{}] 
\exp \left( -\displaystyle\int d^4 x_E L^{(E)}_{0} \right)$
and $J_{E  \mu} (x_E)$ is a four--vector source defined as
\ba
\lefteqn{
J_{E  \mu} (x_E) = J_{E  \mu}^{(1)} (x_E) + J_{E  \mu}^{(2)} (x_E) } 
\nonumber \\
& & = e [ v_{1 \mu} \delta (x_{E 1} \cos\phi - x_{E 4} \sin\phi) 
\delta ({\bf x}_{E t} - {\bf z}_t) 
\nonumber \\
& & + v_{2 \mu} \delta (x_{E 1} \cos\phi + x_{E 4} \sin\phi) 
\delta ({\bf x}_{E t}) ] ~.
\ea
The functional integral
\be
Z^{(E)}_{0} [J_{E }] \equiv 
\displaystyle\int [dA^{(E)}_{}] \exp \left[ -\displaystyle\int d^4 x_E
( L^{(E)}_{0} + i A^{(E)}_{ \mu} J_{E  \mu} ) \right]
\ee
can be evaluated with standard methods (completing the quadratic form in 
the exponent). One thus obtains
\be
Z^{(E)}_{0} [J_{E }] = 
Z'_E \cdot \exp \left[ - {i \over 2} \displaystyle\int d^4 x_E 
\displaystyle\int d^4 y_E 
J_{E  \mu} (x_E) D^{(E)}_{ \mu\nu} (x_E - y_E) J_{E  \nu} (y_E) \right] ~,
\ee
where $D^{(E)}_{ \mu\nu}$ is the free photon Euclidean propagator, i.e.,
\be
D^{(E)}_{ \mu\nu} (x_E - y_E) = \displaystyle\int {d^4 k_E \over (2\pi)^4}
{e^{-ik_E (x_E - y_E)} \over k_E^2}
\left( \delta_{\mu\nu} - (1 - \alpha){k_{E \mu} k_{E \nu} \over k_E^2} 
\right) ~.
\ee
In the following we shall choose the gauge--fixing parameter $\alpha$ equal 
to $1$ ({\it Feynman} gauge). 
From Eqs. (2.22) $\div$ (2.26), we derive the following expression 
for the expectation value (2.21) of the two Euclidean Wilson lines
(including the regulating IR {\it cutoff} in the form of a 
small photon mass $\lambda$, which must be put equal to zero at the end of 
the calculation):
\ba
\lefteqn{
\langle W_{E 1} (\bar{z}_{t}) W_{E 2} (0) \rangle_A \simeq
\langle W_{E 1} (\bar{z}_{t}) \rangle_A \langle W_{E 2} (0) \rangle_A }
\nonumber \\
& & \times \exp \left[ - \displaystyle\int d^4 x_E \displaystyle\int d^4 y_E
J^{(1)}_{E \mu} (x_E) J^{(2)}_{E \mu} (y_E) 
\displaystyle\int {d^4 k_E \over (2\pi)^4}
{e^{-ik_E (x_E - y_E)} \over k_E^2 + \lambda^2} \right] ~.
\ea
Finally, making use of the explicit form (2.23) of the four--vector source 
$J_{E  \mu} (x_E)$ to evaluate the double integral in Eq. (2.27), and using 
also the $O(4)$ {\it plus} translation invariance of the Euclidean theory 
to write:
$\langle W_{E 1} \rangle_A \equiv \langle W_{E 1} (\bar{z}_{t}) \rangle_A
= \langle W_{E 1} (0) \rangle_A = \langle W_{E 2} (0) \rangle_A$,
one finds the result
\be
{ \langle W_{E 1} (\bar{z}_{t}) W_{E 2} (0) \rangle_A \over
\langle W_{E 1} \rangle_A^2  }
\simeq \exp \left[ - e^2 \cot \theta \cdot
\displaystyle\int {d^2 {\bf k}_t \over (2\pi)^2}
{e^{-i{\bf k}_t \cdot {\bf z}_t} \over {\bf k}_t^2 + \lambda^2} \right] 
~.
\ee
We have indicated with $\theta \equiv 2 \phi$ the angle [in the plane 
$(x_1,x_4)$] between the two trajectories in Euclidean four--space. 
The angle $\theta$ in Eq. (2.28) 
is taken in the interval $[0,\pi]$: it is always possible to 
make such a choice by virtue of the $O(4)$ symmetry of the Euclidean theory.
When comparing the two expressions (2.14) and (2.28), we immediately 
recognize that they are linked by the following analytic continuation in 
the angular variables:
\ba
{ \langle W_{E 1} (\bar{z}_{t}) W_{E 2} (0) \rangle_A \over
\langle W_{E 1} \rangle_A^2 }
\mathop{\longrightarrow}_{\theta \to -i \chi}
{ \langle W_1 (z_t) W_2 (0) \rangle_A \over
\langle W_1 \rangle_A^2 } ~;
\nonumber \\
{\rm or:}~~ { \langle W_1 (z_t) W_2 (0) \rangle_A \over \langle W_1
\rangle_A^2 }
\mathop{\longrightarrow}_{\chi \to i \theta}
{ \langle W_{E 1} (\bar{z}_{t}) W_{E 2} (0) \rangle_A \over
\langle W_{E 1} \rangle_A^2 }
~.
\ea
This allows to reconstruct the high--energy scattering amplitude by 
evaluating a correlation of infinite Wilson lines in the Euclidean world, 
then by continuing this quantity in the angular variable, $\theta \to 
-i\chi$, and finally by performing the limit $\chi \to \infty$ (i.e.,
$\beta \to 1$). In fact, from Eq. (2.1) we can write
\ba
\lefteqn{
M_{fi} = \langle \psi_{\alpha}(p'_1) \psi_{\gamma}(p'_2) | M | 
\psi_{\beta}(p_1) \psi_{\delta}(p_2) \rangle } \nonumber \\
& & \mathop{\sim}_{s \to \infty}
-i \cdot 2s \cdot \delta_{\alpha\beta} \delta_{\gamma\delta}
\cdot g_M (t,\chi \to \infty) ~,
\ea
where the quantity $g_M (t,\chi)$ is defined as
\be
g_M (t,\chi) =
\displaystyle\int d^2 {\bf z}_t e^{i {\bf q} \cdot {\bf z}_t}
{ \langle [ W_1 (z_t) - 1 ] [ W_2 (0) - 1 ] \rangle_A \over
\langle W_1 \rangle_A^2 } ~.
\ee
It was shown in Ref. \cite{Nachtmann91} that, in the eikonal approximation,
one can approximate the fermion--field renormalization constant as follows:
\be
Z_\psi \simeq \langle W_1 \rangle_A ~.
\ee
Therefore $g_M (t,\chi \to \infty)$ is exactly the Abelian version of the 
asymptotic amplitude (1.7).
Moreover, whenever $t$ is not exactly equal to zero, i.e.,
${\bf q} \ne 0$ ($t = -{\bf q}^2 <  0$), the expression (2.31) reduces to
\be
g_M (t,\chi) =
\displaystyle\int d^2 {\bf z}_t e^{i {\bf q} \cdot {\bf z}_t}
{ \langle W_1 (z_t) W_2 (0) \rangle_A \over \langle W_1 \rangle_A^2 } ~.
\ee
Therefore
$g(t,\chi)$ turns out to be the Fourier transform, with respect to the 
transverse coordinates ${\bf z}_t$, of the quantity (2.14), which we have 
evaluated in the {\it quenched} approximation.
In the Euclidean theory we can define the corresponding quantity
\be
g_E (t,\theta) =
\displaystyle\int d^2 {\bf z}_t e^{i {\bf q} \cdot {\bf z}_t}
{ \langle W_{E 1} (\bar{z}_{t}) W_{E 2} (0) \rangle_A \over
\langle W_{E 1} \rangle_A^2 } ~.
\ee
This is just
the Fourier transform, with respect to the transverse coordinates ${\bf z}_t$, 
of the quantity (2.28), which we have evaluated in the {\it quenched} 
approximation. Using the relation (2.29), we can derive that
\ba
g_E (t,\theta)
\mathop{\longrightarrow}_{\theta \to -i \chi}
g_E(t,-i \chi) = g_M (t,\chi) ~;
\nonumber \\
{\rm or:}~~ g_M (t,\chi)
\mathop{\longrightarrow}_{\chi \to i \theta}
g_M (t,i\theta) = g_E (t,\theta)
~.
\ea

\newsection{The case of QCD at order $O(g_R^4)$}

\noindent
In this section we shall see that the same relationship of analytic
continuation appears to be valid also for the case of a non--Abelian
gauge theory: we shall prove this up to the order $O(g_R^4)$ in
perturbation theory ($g_R$ being the renormalized coupling constant).
Let us consider the following quantity, defined in
Minkowski space--time:
\be
g_M (t,p_1 \cdot p_2) = {1 \over Z_W^2}
\displaystyle\int d^2 {\bf z}_t e^{i {\bf q} \cdot {\bf z}_t}
\langle [ W_1 (z_t) - {\bf 1} ]_{ij} [ W_2 (0) - {\bf 1} ]_{kl} \rangle_A ~,
\ee
where $p_1$ and $p_2$ are the four--momenta [lying (for example) in the plane 
$(x^0,x^1)$], which define the two Wilson lines $W_1$ and $W_2$. 
By virtue of the Lorentz symmetry, we can define $p_1$ and $p_2$ as 
in Eq. (2.4): that is, we choose, as the reference frame, the c.m.s. of the two 
particles, moving with speed $\beta$ and $-\beta$ along the $x^1$--direction.
Of course, $g_M$ can only depend on the scalar quantities constructed with 
the vectors $p_1$, $p_2$ and $q = (0,0,{\bf q})$: the only possibilities 
are $q^2 = -{\bf q}^2 = t$ and $p_1 \cdot p_2$, because $p_1 \cdot q =
p_2 \cdot q = 0$ and $p_1^2 = p_2^2 = m^2$ are fixed.
In such a reference frame, we can write $p_1 \cdot p_2 = m^2 \cosh \chi$,
where $\chi = 2\psi$ (with $\beta = \tanh \psi$) is the hyperbolic angle
[in the plane $(x^0,x^1)$] between the two Wilson lines $W_1$ and $W_2$.
The Wilson lines are defined as
\ba
W_1 (z_t) &\equiv&
{P} \exp \left[ -ig \displaystyle\int_{-\infty}^{+\infty}
A_\mu (z_t + {p_1 \over m} \tau) {p_1^\mu \over m} d\tau \right] ~;
\nonumber \\
W_2 (0) &\equiv&
{P} \exp \left[ -ig \displaystyle\int_{-\infty}^{+\infty}
A_\mu ({p_2 \over m} \tau) {p_2^\mu \over m} d\tau \right] ~,
\ea
Moreover, we have put, in Eq. (3.1),
\be
Z_W \equiv {1 \over N_c} \langle \Tr [ W_1 (z_t) ] \rangle_A
= {1 \over N_c} \langle \Tr [ W_1 (0) ] \rangle_A = {1 \over N_c}
\langle \Tr [ W_2 (0) ] \rangle_A ~.
\ee
(The two last equalities come from the Poincar\'e invariance.)
This is a sort of Wilson--line's renormalization constant:
as shown in Ref. \cite{Nachtmann91}, $Z_W$ coincides with the fermion 
renormalization constant $Z_\psi$ in the eikonal approximation.
We want to explicitly evaluate the quantity (3.1) up to the order $O(g_R^4)$ in 
perturbation theory.
This corresponds to evaluate the Feynman diagrams in Figs. 2 and 3, 
where the two horizontal oriented lines 
represent the Wilson lines $W_1$ and $W_2$.
First of all we need to expand $Z_W$ up to the
order $O(g_R^2)$ in perturbation theory:
\be
Z_W = 1 + Z_W^{(2)} g_R^2 + O(g_R^4) ~.
\ee
Clearly, we do not need to consider also the $O(g_R^4)$ piece, of the form
$Z_W^{(4)} g_R^4$, since the expectation value 
$\langle \ldots \rangle_A$ in Eq. (3.1) is an 
object of order $O(g_R^2)$. As will become clear in the following, we do not 
need to know the explicit expression for the coefficient $Z_W^{(2)}$.
Since we are 
interested in evaluating the quantity (3.1) up to the order $O(g_R^4)$, we need 
to consider also the effects of the renormalizations of the fields and the 
coupling constant $g$, up to the order $O(g_R^2)$. Using the conventional 
notation, we write
\be
A_\mu^a = Z_3^{1/2} A_{R \mu}^a ~~ ; ~~ g = Z_g g_R ~,
\ee
where the suffix ``$R$'' denotes the renormalized quantities. Therefore, we 
have that
\be
W_i (z_t) =
{P} \exp \left[ -i Z_{1 W} g_R \displaystyle\int_{-\infty}^{+\infty}
A_{R \mu} (z_t + {p_1 \over m} \tau) {p_1^\mu \over m} d\tau \right] ~,
\ee
where the renormalization constant $Z_{1 W}$ is defined as
\be
Z_{1 W} \equiv  Z_g Z_3^{1/2} = 1 + Z_{1 W}^{(2)} g_R^2 + O(g_R^4) ~.
\ee
Since we are interested in evaluating the amplitude
\be
M (t,\chi) = 
\displaystyle\int d^2 {\bf z}_t e^{i {\bf q} \cdot {\bf z}_t}
\langle [ W_1 (z_t) - {\bf 1} ]_{ij} [ W_2 (0) - {\bf 1} ]_{kl} \rangle_A ~,
\ee
up to the order $O(g_R^4)$, the effects of $Z_{1 W}$ are visible when we 
expand the Wilson lines $W_i$ only up to the first order in $g$:
\be
W_i (z_t) = {\bf 1} -i Z_{1 W} g_R \displaystyle\int_{-\infty}^{+\infty}
A_{R \mu} (z_t + {p_1 \over m} \tau) {p_1^\mu \over m} d\tau +  \ldots ~.
\ee
This corresponds to consider only the diagrams of the one--gluon--exchanged 
type, having the following amplitude:
\be
M_{(1,1)} = Z_{1 W}^2 M_{R(1,1)} = M_{R(1,1)} + (Z_{1 W}^2 - 1) M_{R(1,1)} ~,
\ee
where $M_{R(1,1)}$ is the ``renormalized'' one--gluon--exchanged amplitude,
obtained from $M_{(1,1)}$ by substituting the coupling constant $g$ in 
front and the gluon field $A^\mu$ with the corresponding renormalized 
quantities:
\be
M_{R(1,1)} = -g_R^2 (T^a)_{ij} (T^b)_{kl} {p_1^\mu p_2^\nu \over m^2}
\displaystyle\int d^2 {\bf z}_t e^{i {\bf q} \cdot {\bf z}_t}
\displaystyle\int d\tau \displaystyle\int d\omega 
\langle A_{R \mu}^a (z_t + {p_1 \over m} \tau) 
A_{R \nu}^b ({p_2 \over m} \omega) \rangle_A ~.
\ee
In our notation, $M_{(i,j)}$
denotes the contribution to the amplitude M, defined in Eq. (3.8), obtained 
after expanding the Wilson lines $W_1$ up to the order $O(g^i)$ (i.e., up to 
the term containing $i$ gluon fields) and expanding the other Wilson line 
$W_2$ up to the order $O(g^j)$ (i.e., up to the term containing $j$ gluon 
fields). Moreover, we define: $M_{R (i,j)} \equiv Z_{1 W}^{-(i+j)} 
M_{(i,j)}$. If one wants to compute $M_{(1,1)}$ up to the order $O(g_R^4)$, 
one must proceed as follows, by virtue of Eqs. (3.10) and (3.4):
\be
M_{(1,1)} |_{g_R^4} = M_{R(1,1)} |_{g_R^4} + 2 Z_{1 W}^{(2)} g^2_R
\cdot M_{R(1,1)} |_{g^2_R} ~.
\ee
The expression for $M_{R(1,1)} |{g^2_R}$ can be immediately derived: it 
corresponds to the diagram shown in Fig. 2(a). In a given Lorentz gauge
with a (bare) gauge parameter $\alpha$, the free gluon--field propagator is 
given by
\be
G_{\mu \nu}^{a b} (x-y) =
-i \delta_{a b} \displaystyle\int{d^4 k \over (2 \pi)^4}
{1 \over k^2 + i \varepsilon} 
\left[ g_{\mu\nu} - (1 - \alpha) { k_\mu k_\nu \over k^2 + i\varepsilon}
\right] e^{-ik (x-y)} ~.
\ee
One thus finds that
\be
M^{(a)} (t,\chi) = M_{R(1,1)} (t,\chi) |_{g^2_R} = 
g^2_R {1 \over t} i \coth \chi \cdot (G_1)_{ij,kl} ~,
\ee
where $G_1$ is the color factor for the one--gluon--exchanged process:
\ba
G_1 &\equiv& T^c_{(1)} \otimes T^c_{(2)} ~, \nonumber \\
(G_1)_{ij,kl} &\equiv& (T^c)_{ij} (T^c)_{kl} ~.
\ea
Let us observe that $M_{R (1,1)} |_{g_R^2}$ is gauge--independent, since 
the gauge parameter $\alpha$ does not appear at the right--hand--side of
Eq. (3.14).
The last term of Eq. (3.12) can be represented by the diagrams in Figs.
3(q) and 3(r), in which we have put a counterterm of the form 
$-i Z_{1 W}^{(2)} g_R^3 (T^a)_{ij}$ in one of the two vertices between the 
gluon line and a Wilson line. So we have that
\ba
\lefteqn{
M^{(q)} (t,\chi) = M^{(r)} (t,\chi) }
\nonumber \\
& & = Z_{1 W}^{(2)} g_R^2 \cdot
M_{R(1,1)} (t,\chi) |_{g^2_R} = Z_{1 W}^{(2)} g_R^4 \cdot
{1 \over t} i \coth \chi \cdot (G_1)_{ij,kl} ~.
\ea
The expression for the one--gluon--exchanged renormalized amplitude up to 
the order $O(g_R^4)$, i.e., $M_{R(1,1)} |_{g_R^4}$, is given by the sum of 
the contributions from the diagrams shown in Figs. 2(a), 3(l) to 3(p):
this last one represents the insertion of a counterterm 
$(Z_3 - 1) \delta_{ab} (k^\mu k^\nu - g^{\mu\nu} k^2)$ into the 
gluon line. In other words, one has to compute the quantity (3.11), using the 
renormalized gluon propagator up to the order $O(g^2_R)$ when evaluating 
the expectation value $\langle A_{R \mu}^a (z_t + {p_1 \over m} \tau) 
A_{R \nu}^b ({p_2 \over m} \omega) \rangle_A$. 
That is, in a given Lorentz gauge 
with a renormalized gauge parameter $\alpha_R = Z_3^{-1} \alpha$:
\be
\langle A_{R \mu}^a (x) A_{R \nu}^b (y) \rangle_A =
-i \displaystyle\int {d^4 k \over (2\pi)^4} e^{-ik(x-y)}
\tilde{D}^{ab}_{R \mu \nu} (k) ~,
\ee
where $\tilde{D}^{ab}_{R \mu \nu} (k)$ is given by
\be
\tilde{D}^{ab}_{R \mu \nu} (k) =  Z_3^{-1} \tilde{D}^{ab}_{\mu \nu} (k) 
= {\delta_{ab} \over k^2 + i\varepsilon} \left[
{ g_{\mu \nu} - {k_\mu k_\nu \over k^2 + i\varepsilon} \over
1 + \Pi_R (k^2) } + \alpha_R {k_\mu k_\nu \over k^2 + i\varepsilon} 
\right] ~.
\ee
$\Pi_R (k^2)$ is a finite function of order $O(g_R^2)$, whose precise 
form depends on the renormalization scheme which has been adopted:
\be
\Pi_R (k^2) = g_R^2 F^{(2)} (k^2) + O(g_R^4) ~.
\ee
At this point one can derive the full expression for the amplitude
$M_{R(1,1)}$:
\be
M_{R(1,1)} (t,\chi) = { M_{R(1,1)} (t,\chi) |_{g^2_R} \over 1 + \Pi_R (t) } =
g^2_R {1 \over t [ 1 + \Pi_R (t) ]} i \coth \chi \cdot 
(G_1)_{ij,kl} ~.
\ee
Let us observe that, differently from $M_{R(1,1)} |_{g_R^2}$, 
the value of $M_{R(1,1)}$ is gauge--dependent, since 
the gauge parameter $\alpha_R$ does appear inside $\Pi_R$ at the 
right--hand--side of Eq. (3.20). This is also generally true for the other 
diagrams in Figs. (2) and (3). Therefore, for the following calculations, 
we shall fix the gauge parameter $\alpha_R$ to $1$ (the so--called 
{\it Feynman} gauge).
Eq. (3.20) is the full expression for $M_{R(1,1)}$, not truncated at any 
perturbative order. Yet, we are only interested in the espression for
$M_{R(1,1)}$ up to the order $O(g_R^4)$:
\be
M_{R(1,1)} (t,\chi) |_{g_R^4} = 
g^2_R {1 \over t} [ 1 - g_R^2 F^{(2)} (t) ] i \coth \chi \cdot 
(G_1)_{ij,kl} ~.
\ee
Therefore, the contribution coming from the 
$O(g_R^4)$ diagrams shown in Figs. 3(l) to 3(p) is given by
\be
M^{(l)} (t,\chi) + \ldots + M^{(p)} (t,\chi) =
-g_R^4 { F^{(2)} (t) \over t } i \coth \chi \cdot (G_1)_{ij,kl} ~.
\ee
The contribution of order $O(g_R^4)$ coming from the two Feynman diagrams shown 
in Figs. 2(b) and 2(c) is obtained by multiplying the two 
pieces of order $O(g^2)$ for each of the two terms $W_i - {\bf 1}$ in
Eq. (3.8): i.e., we must evaluate the contribution $M_{(2,2)} |_{g_R^4}$.
The contribution $M^{(b)}$ coming from the graph in Fig. 2(b) is 
conventionally called {\it ladder} term, while the other contribution
$M^{(c)}$, coming from the graph in Fig. 2(c), will be called 
{\it cross} term.
These two contributions can be evaluated in perturbation theory and the 
final result is (in the {\it Feynman} gauge, where $\alpha_R = 1$)
\ba
\lefteqn{
M^{(b)} (t,\chi) + M^{(c)} (t,\chi) = }
\nonumber \\
& & = M_{(2,2)} (t,\chi) |_{g_R^4} = 
M^{(G_1)} (t,\chi) \cdot (G_1)_{ij,kl} +
M^{(G_2)} (t,\chi) \cdot (G_2)_{ij,kl} ~,
\ea
where $G_1$ has been already defined in Eq. (3.15) and $G_2$ is the color 
factor for the {\it ladder} process in Fig. 2(b), i.e.,
\ba
G_2 &\equiv& (T^a_{(1)} T^b_{(1)}) \otimes (T^a_{(2)} T^b_{(2)}) ~,
\nonumber \\
(G_2)_{ij,kl} &\equiv& (T^a T^b)_{ij} (T^a T^b)_{kl} ~.
\ea
In writing Eq. (3.23), we have made use of the following relation for the 
color factors:
\ba
\lefteqn{
(T^a T^b)_{ij} (T^b T^a)_{kl} } \nonumber \\
& & = (T^a T^b)_{ij} (T^a T^b)_{kl} + {N_c \over 2} (T^c)_{ij} (T^c)_{kl} 
\equiv (G_2)_{ij,kl} + {N_c \over 2} (G_1)_{ij,kl} ~,
\ea
which can be easily recovered using the algebra of the generators $T^a$,
i.e., $[ T^a , T^b ] = i f_{abc} T^c$; $N_c$ is the number of colors of the 
theory (the gauge group is $SU(N_c)$).
The coefficients $M^{(G_1)} (t,\chi)$ and $M^{(G_2)} (t,\chi)$
in front of the color factors in Eq. (3.23) are found to be
\ba
M^{(G_1)} (t,\chi) &=& i {N_c g_R^4 \over 4 \pi} I(t)
\chi \coth^2 \chi ~; \nonumber \\
M^{(G_2)} (t,\chi) &=& -{1 \over 2} g_R^4 I(t)
\coth^2 \chi ~.
\ea
In the previous expression we have adopted the conventional notation
\ba
\lefteqn{
I(t) = \displaystyle\int {d^2 {\bf k}_t \over (2\pi)^2} 
{1 \over {\bf k}_t^2 + \lambda^2}
{1 \over ({\bf q} - {\bf k}_t)^2 + \lambda^2} } \nonumber \\
& & = \displaystyle\int d^2 {\bf z}_t 
e^{i{\bf q} \cdot {\bf z}_t} \left(
{d^2 {\bf k}_t \over (2\pi)^2} 
{ e^{-i{\bf k}_t \cdot {\bf z}_t} \over 
{\bf k}_t^2 + \lambda^2} \right)^2 ~.
\ea
(Remember that: $t = -{\bf q}^2$). The quantity $\lambda$ is the usual 
regularizing gluon mass, used as an IR {\it cutoff}: it must be put equal 
to zero at the end of the calculation.

We shall now compute the contribution from the diagrams in Figs. 2(d) 
to 2(i). They are $O(g_R^4)$ diagrams, obtained after expanding one of 
the two Wilson lines up to the order $O(g^3)$, and the remaining one up
the first order in $g$. We shall denote the sum of all these contributions by 
$M_{(3,1)} |_{g_R^4} + M_{(1,3)} |_{g_R^4}$, in agreement with the notation 
introduced above. One thus finds that
\ba
M_{(3,1)} |_{g_R^4} &=&
Z_W^{(2)} g_R^2 \cdot M_{R(1,1)} |_{g_R^2} + \Delta M_{(3,1)} ~,
\nonumber \\
M_{(1,3)} |_{g_R^4} &=&
Z_W^{(2)} g_R^2 \cdot M_{R(1,1)} |_{g_R^2} + \Delta M_{(1,3)} ~,
\ea
where $\Delta M_{(3,1)}$ and $\Delta M_{(1,3)}$ are divergent quantities, 
whose regularized 
expressions depend on the adopted renormalization scheme. In the 
{\it minimal subtraction} (MS) renormalization scheme one finds that
\be
\Delta M_{(3,1)} = \Delta M_{(1,3)} = 
M_{R(1,1)} |_{g^2_R} \cdot {g^2_R \over (4\pi)^2}
N_c \left[ {1 \over \varepsilon} + B \right] ~,
\ee
where $\varepsilon = (4 - D)/2$, $D$ being the number of space--time dimensions,
and $B$ is a finite number (as $\varepsilon$ goes to zero).
In the same renormalization scheme (MS), one also has that (always in the 
{\it Feynman} gauge $\alpha_R = 1$):
\be
Z_{1 W} = 1 + Z_{1 W}^{(2)} g_R^2 + O(g_R^4) ~,~ {\rm with:} ~~
Z_{1 W}^{(2)} = - {g_R^2 \over (4\pi)^2} N_c {1 \over \varepsilon} ~.
\ee
Therefore, from Eqs. (3.29) and (3.30), one immediately concludes that
\be
\Delta M_{(3,1)} + \Delta M_{(1,3)} + 2 Z_{1 W}^{(2)} g_R^2 \cdot 
M_{R(1,1)} |_{g_R^2} = M_{R(1,1)} |_{g^2_R} \cdot {g^2_R \over (4\pi)^2} 
2 N_c B ~.
\ee
In other words, the divergence contained in $\Delta M_{(3,1)} + 
\Delta M_{(1,3)}$ is exactly 
calcelled out by the two diagrams with the counterterm $Z_{1 W}^{(2)}$,
represented in Figs. 3(q) and 3(r).
Finally, we have to evaluate the two diagrams in Figs. 3(j) and 3(k): in 
agreement with the notation introduced above, we shall denote their 
contribution by $M_{(2,1)} |_{g_R^4}$ and $M_{(1,2)} |_{g_R^4}$,
respectively.
However, explicit calculations show that their contribution vanishes:
\be
M_{(2,1)} |_{g_R^4} = M_{(1,2)} |_{g_R^4} = 0 ~.
\ee
At this point we can sum up all the contributions previously evaluated in 
order to find the complete expression for the amplitude $M$, defined by
Eq. (3.8), up to the order $O(g_R^4)$. We find that
\be
M (t,\chi) |_{g_R^4} =
\left[ 1 + \left( 2 Z_W^{(2)}
- F^{(2)} (t) + {2 N_c B  \over (4\pi)^2} \right) g^2_R \right] 
\cdot M_{R(1,1)} (t,\chi) |_{g^2_R} + M_{R(2,2)} (t,\chi) ~.
\ee
Introducing here the expressions found above for $M_{R(1,1)} |_{g_R^2}$ 
[see Eq. (3.14)] and for $M_{R(2,2)} |_{g_R^4}$ [see Eqs. (3.23) and (3.26)], 
we finally find the following expression for $M (t,\chi) |_{g_R^4}$:
\ba
\lefteqn{
M (t,\chi) |_{g_R^4} = g^2_R {1 \over t} i \coth \chi }
\nonumber \\
& & \times \left[ 1 + \left( 2 Z_W^{(2)} - F^{(2)} (t) + 
{2 N_c B \over (4\pi)^2} + {N_c \over 4\pi} t I(t) \chi \coth \chi
\right) g^2_R \right] \cdot (G_1)_{ij,kl}
\nonumber \\
& & -{1 \over 2} g_R^4 I(t) \coth^2 \chi \cdot (G_2)_{ij,kl} ~.
\ea
This expression allows us to immediately derive the quantity $g_M (t,\chi)$,
defined by Eq. (3.1), up to the order $O(g_R^4)$. In fact, making use also of
the expansion (3.30) for the renormalization constant $Z_{1 W}$, one finds that:
\ba
\lefteqn{
g_M (t,\chi) |_{g_R^4} = { M (t,\chi) \over Z_W^2 } 
\displaystyle\vert_{g_R^4} =
M (t,\chi) |_{g_R^4} - 2 Z_W^{(2)} g_R^2 
\cdot M (t,\chi) |_{g^2_R} }
\nonumber \\
& & = g^2_R {1 \over t} i \coth \chi
\left[ 1 - \left( F^{(2)} (t) + {2 N_c B \over (4\pi)^2} + 
{N_c \over 4\pi} t I(t) \chi \coth \chi \right) g^2_R \right] 
\cdot (G_1)_{ij,kl}
\nonumber \\
& & -{1 \over 2} g_R^4 I(t) \coth^2 \chi \cdot (G_2)_{ij,kl} ~.
\ea
The quark--quark scattering amplitude in the high--energy limit turns out
to be, up to the order $O(g_R^4)$,
\ba
\lefteqn{
f_{M} (s,t) |_{g_R^4} \mathop{\sim}_{s \to \infty}
-i \cdot \delta_{\alpha\beta} \delta_{\gamma\delta}
\cdot 2s \cdot g_M (t, \chi \to \infty) |_{g_R^4} }
\nonumber \\
& & = \delta_{\alpha\beta} \delta_{\gamma\delta} \left[
g_R^2 {2s \over t} 
\left[ 1 - \bar{\alpha}(t) \ln s \right]
\cdot (G_1)_{ij,kl} + i g_R^4 s I(t) \cdot (G_2)_{ij,kl} \right] ~,
\ea
where we have used the notation
\be
\bar{\alpha} (t) \equiv -{N_c g_R^2 \over 4 \pi} t I(t)
= {N_c g_R^2 \over 4 \pi} {\bf q}^2 I(t) ~.
\ee
We have used the fact that both $\beta$ and $\psi$ (or equivalently $\chi$) 
are dependent on $s$. In fact, from $E = m/\sqrt{1 - \beta^2}$ and from
$s = 4E^2$, one immediately finds that
\be
\beta = \sqrt{ 1 - {4 m^2 \over s} } ~.
\ee
By inverting this equation and using the relation $\beta = \tanh \psi$,
we derive that
\be
s = 4 m^2 \cosh^2 \psi = 2 m^2 ( \cosh \chi + 1 ) ~.
\ee
Therefore, in the high--energy limit $s \to 
\infty$ (or $\beta \to 1$), the hyperbolic angle $\chi = 2\psi$ is 
essentially equal to the logarithm of $s$:
\be
\chi = 2\psi \mathop{\sim}_{s \to \infty} \ln s ~.
\ee
Moreover, $\coth \chi \sim 1$ in this limit. 
This is why we have been able to approximate the $O(g_R^4)$ 
term which multiplies $(G_1)_{ij,kl}$ as reported in Eq. (3.36).
The result (3.36) is exactly what can be found by applying 
ordinary perturbation theory to evaluate the scattering amplitude up to the 
order $O(g_R^4)$ \cite{Cheng-Wu-book} \cite{Lipatov}.
In particular, as was pointed out in the Introduction, the $\ln s$ 
factor in Eq. (3.36) comes from the fact that it is really a singular limit 
to take 
the Wilson lines in (3.1) exactly on the light cone. As first predicted in 
\cite{Verlinde}, a proper regularization of these singularities give rise 
to the $\ln s$ dependence of the amplitude, as confirmed by the experiments 
on hadron--hadron scattering processes.

We want now to repeat the analogous calculation for the 
Euclidean theory. Let us consider, therefore, the following quantity, defined 
in Euclidean space--time:
\be
g_E (t,v_1 \cdot v_2) = {1 \over Z_{E W}^2}
\displaystyle\int d^2 {\bf z}_t e^{i {\bf q} \cdot {\bf z}_t}
\langle [ W_{E 1} (\bar{z}_{t}) - {\bf 1} ]_{ij} [ W_{E 2} (0) - {\bf 1} ]_{kl} 
\rangle_A ~,
\ee
where $\bar{z}_{t} = (0, {\bf z}_t, 0)$ and
the expectation value $\langle \ldots \rangle_A$ must be intended now as a 
functional integration with respect to the gauge variable $A^{(E)}_{ \mu}$ in 
the Euclidean theory.
The Euclidean four--vectors $v_1$ and $v_2$ [lying (for example) in the plane 
$(x_1,x_4)$] 
define the two Wilson lines $W_{E 1}$ and $W_{E 2}$: we can take $v_1$ and 
$v_2$ normalized to $1$, with respect to the Euclidean scalar product
(that is, $v_1^2 = v_2^2 = 1$). Clearly,
$g_E$ can only depend on the scalar variables constructed using the 
vectors $v_1$, $v_2$ and $q_E = (0,{\bf q},0)$: they are $q_E^2 = {\bf q}^2 
= -t$ and $v_1 \cdot v_2$, since $q_E \cdot v_1 = q_E \cdot v_2 = 0$ and
$v_1^2 = v_2^2 = 1$ are fixed.
By virtue of the $O(4)$ symmetry of the Euclidean theory, we can choose
a reference frame in which $v_1$ and $v_2$ have the following values:
\ba
v_1 &=& (\sin \phi, {\bf 0}_t, \cos \phi ) ~; \nonumber \\
v_2 &=& (-\sin \phi, {\bf 0}_t, \cos \phi ) ~,
\ea
with a value of $\phi$ between $0$ and $\pi / 2$ (so that the angle
$2 \phi$ between the two trajectories is in the interval $[0,\pi]$).
In such a reference frame, we can write $v_1 \cdot v_2 = \cos \theta$, 
where $\theta = 2\phi$ is the angle [in the plane $(x_1,x_4)$] between the 
two Euclidean Wilson lines $W_{E 1}$ and $W_{E 2}$.
These last are defined as
\ba
W_{E 1} (\bar{z}_{t}) &\equiv&
{P} \exp \left[ -ig \displaystyle\int_{-\infty}^{+\infty}
A^{(E)}_{ \mu} (\bar{z}_{t} + v_1 \tau) v_{1 \mu} d\tau \right] ~;
\nonumber \\
W_{E 2} (0) &\equiv&
{P} \exp \left[ -ig \displaystyle\int_{-\infty}^{+\infty}
A^{(E)}_{ \mu} (v_2 \tau) v_{2 \mu} d\tau \right] ~,
\ea
where $A^{(E)}_{ \mu} = A^{(E) a}_{ \mu} T^a$. Moreover, we have put
\be
Z_{E W} \equiv {1 \over N_c} \langle \Tr [ W_{E 1} (\bar{z}_{t}) ] \rangle
= {1 \over N_c} \langle \Tr [ W_{E 1} (0) ] \rangle = {1 \over N_c}
\langle \Tr [ W_{E 2} (0) ] \rangle ~.
\ee
(The two last equalities come from the $O(4)$ {\it plus} 
translation invariance.)
We want to explicitly evaluate the quantity (3.41) up to the order 
$O(g_R^4)$ in 
perturbation theory. Therefore, we have to evaluate the 
Feynman diagrams in Figs. 2 and 3, where the two horizontal oriented lines 
now represent the Euclidean Wilson lines $W_{E 1}$ and $W_{E 2}$.
As before, we need to expand $Z_{E W}$ only up to the
order $O(g_R^2)$ in perturbation theory:
\be
Z_{E W} = 1 + Z_{E W}^{(2)} g_R^2 + O(g_R^4) ~,
\ee
even if, as will become clear in the following, we shall not 
need to know the explicit expression for the coefficient $Z_{E W}^{(2)}$.
As in the previous case, we need 
to consider also the effect of the renormalizations of the fields and the 
coupling constant $g$, up to the order $O(g_R^2)$, that is:
\be
A^{(E) a}_{ \mu} = Z_3^{1/2} A^{(E) a}_{ R \mu} ~~ ; ~~ g = Z_g g_R ~,
\ee
where the suffix ``$R$'' denotes the renormalized quantities. Therefore we 
have that
\be
W_{E i} (\bar{z}_{t}) =
{P} \exp \left[ -i Z_{1 W} g_R \displaystyle\int_{-\infty}^{+\infty}
A^{(E)}_{R \mu} (\bar{z}_{t} + v_1 \tau) v_{1 \mu} d\tau \right] ~,
\ee
the renormalization constant $Z_{1 W}$ being defined by Eq. (3.7).
The renormalization constants in Eq. (3.46) are the same as those in 
Eq. (3.5) for the Minkowski world, if we adopt the same renormalization scheme
(for example, the MS scheme) for the Euclidean theory and the Minkowskian one.
In such a case, the correspondence
\be
A_0 (x) \rightarrow i A^{(E)}_{ 4}  (x_E) ~, ~~
A_k (x) \rightarrow A^{(E)}_{ k} (x_E) ~~~ 
({\rm with:}~~x^0 \rightarrow -i x_{E 4})
\ee
between the bare gluon fields in the two theories, turns into the same 
correspondence for the renormalized gluon fields:
\be
A_{R 0} (x) \rightarrow i A^{(E)}_{R 4}  (x_E) ~, ~~
A_{R k} (x) \rightarrow A^{(E)}_{R k} (x_E) ~~~ 
({\rm with:}~~x^0 \rightarrow -i x_{E 4}) ~.
\ee
[As a matter of fact, the renormalization constants $Z_3$, $Z_g$, etc., are
always evaluated in the Euclidean world, also when they refer to the
Minkowskian one: when evaluating these constants in
Minkowski (momentum) four--space, one always performs a Wick rotation to
Euclidean (momentum) four--space.]

We shall evaluate the amplitude
\be
E (t,\theta) = 
\displaystyle\int d^2 {\bf z}_t e^{i {\bf q} \cdot {\bf z}_t}
\langle [ W_{E 1} (\bar{z}_{t}) - {\bf 1} ]_{ij} [ W_{E 2} (0) - {\bf 1} ]_{kl} 
\rangle_A ~,
\ee
up to the order $O(g_R^4)$. As in the previous case, the effects of $Z_{1 W}$ 
are visible when we consider only the diagrams of the one--gluon--exchanged 
type, having the following amplitude:
\be
E_{(1,1)} = Z_{1 W}^2 E_{R(1,1)} = E_{R(1,1)} + (Z_{1 W}^2 - 1) E_{R(1,1)} ~,
\ee
$E_{R(1,1)}$ being the ``renormalized'' one--gluon--exchanged amplitude:
\be
E_{R(1,1)} = -g_R^2 (T^a)_{ij} (T^b)_{kl} v_{1 \mu} v_{2 \nu}
\displaystyle\int d^2 {\bf z}_t e^{i {\bf q} \cdot {\bf z}_t}
\displaystyle\int d\tau \displaystyle\int d\omega 
\langle A^{(E) a}_{R \mu} (\bar{z}_{t} + v_1 \tau) 
A^{(E) b}_{R \nu} (v_2 \omega) \rangle_A ~.
\ee
In our notation, $E_{(i,j)}$
denotes the contribution to the amplitude E, defined in Eq. (3.50), obtained 
after expanding the Euclidean Wilson line $W_{E 1}$ up to the order $O(g^i)$ 
(i.e., up to the term containing $i$ gluon fields) and expanding the other 
Euclidean Wilson line $W_{E 2}$ up 
to the order $O(g^j)$ (i.e., up to the term containing $j$ gluon fields).
Moreover, we define $E_{R (i,j)} \equiv Z_{1 W}^{-(i+j)} E_{(i,j)}$.
We have to compute $E_{(1,1)}$ up to the order $O(g_R^4)$, which, by virtue 
of Eqs. (3.51) and (3.7), is given by
\be
E_{(1,1)} |_{g_R^4} = E_{R(1,1)} |_{g_R^4} + 2 Z_{1 W}^{(2)} g^2_R 
\cdot E_{R(1,1)} |_{g^2_R} ~.
\ee
The expression for $E_{R(1,1)} |{g^2_R}$, corresponding to the diagram shown 
in Fig. 2(a), can be derived using in the calculation
the Euclidean free gluon--field propagator. This last, in any given 
Lorentz gauge with a (bare) gauge parameter $\alpha$, is given by
\be
G^{(E) ab}_{ \mu \nu} (x_E - y_E) =
\delta_{a b} \displaystyle\int{d^4 k_E \over (2 \pi)^4}
{1 \over k_E^2} 
\left[ \delta_{\mu\nu} - (1 - \alpha) {k_{E \mu} k_{E \nu} \over k_E^2}
\right] e^{-ik_E (x_E - y_E)} ~.
\ee
The contribution coming from the
one--gluon--exchange process pictorially represented in Fig. 2(a),
comes out to be, with the notation already introduced for the color 
factor,
\be
E^{(a)} (t,\theta) = E_{R(1,1)} (t,\theta) |_{g^2_R}
= g_R^2 {1 \over t} \cot \theta \cdot (G_1)_{ij,kl} ~.
\ee
The last term of Eq. (3.53), represented by the diagrams in Figs. 3(q) and 
3(r), is given by
\be
E^{(q)} (t,\theta) = E^{(r)} (t,\theta) = Z_{1 W}^{(2)} g_R^2 \cdot
E_{R(1,1)} (t,\theta) |_{g^2_R} = Z_{1 W}^{(2)} g_R^4 \cdot
{1 \over t} \cot \theta \cdot (G_1)_{ij,kl} ~.
\ee
The first term in Eq. (3.53), i.e., $E_{R(1,1)} |_{g_R^4}$, is the 
one--gluon--exchanged renormalized amplitude up to the order $O(g_R^4)$.
It is given by the sum of 
the contributions from the diagrams shown in Figs. 2(a), 3(l) to 3(p).
[This last one represents the insertion of a counterterm 
$(Z_3 - 1) \delta_{ab} (k_{E \mu} k_{E \nu} - \delta_{\mu\nu} k_E^2)$ into the 
gluon line.] 
Therefore, one has to compute the quantity (3.52) using the 
renormalized gluon propagator up to the order $O(g^2_R)$ when evaluating 
the expectation value $\langle A^{(E) a}_{ R \mu} (\bar{z}_{t} + v_1 \tau) 
A^{(E) b}_{ R \nu} (v_2 \omega) \rangle_A$. That is:
\be
\langle A^{(E) a}_{ R \mu} (x_E) A^{(E) b}_{ R \nu} (y_E) \rangle_A =
\displaystyle\int {d^4 k_E \over (2\pi)^4} e^{-ik_E (x_E - y_E)}
\tilde{D}^{(E) ab}_{ R \mu \nu} (k_E) ~.
\ee
The expression for $\tilde{D}^{(E) ab}_{ R \mu \nu} (k_E)$ can be 
derived from the corresponding expression (3.18) for
$\tilde{D}^{ab}_{R \mu \nu} (k)$, making use of the 
correspondence law (3.49) between the (renormalized) gluon field in
Minkowski four--space and the (renormalized) gluon field in Euclidean 
four--space. $\tilde{D}^{(E) ab}_{ R \mu \nu} (k_E)$ is obtained from
$\tilde{D}^{ab}_{R \mu \nu} (k)$ by making the replacements
\ba
k^2 \rightarrow -k_E^2 ~~ &({\rm i.e.,}& ~~ k^0 \rightarrow i k_{E 4} ~, ~~
{\bf k} \rightarrow {\bf k}_E) ~; \nonumber \\
g_{\mu\nu} \rightarrow -\delta_{\mu\nu} ~~ &;& ~~
k_\mu k_\nu \rightarrow k_{E \mu} k_{E \nu} ~.
\ea
Therefore, in a Lorentz gauge with a renormalized gauge parameter 
$\alpha_R = Z_3^{-1} \alpha$, $\tilde{D}^{(E) ab}_{R \mu \nu} (k_E)$ 
is given by
\be
\tilde{D}^{(E) ab}_{ R \mu \nu} (k_E) =  
Z_3^{-1} \tilde{D}^{(E) ab}_{ \mu \nu} (k_E) = 
{\delta_{ab} \over k_E^2} \left[
{ \delta_{\mu \nu} - {k_{E \mu} k_{E \nu} \over k_E^2} \over
1 + \Pi_R (-k_E^2) } + \alpha_R {k_{E \mu} k_{E \nu} \over k_E^2}
\right] ~,
\ee
where $\Pi_R$ is exactly the same finite function of order $O(g_R^2)$, 
appearing in the expression (3.18) for the gluon propagator in Minkowski 
space--time:
\be
\Pi_R (-k_E^2) = g_R^2 F^{(2)} (-k_E^2) + O(g_R^4) ~.
\ee
As said before, the precise form of $\Pi_R$ depends on the renormalization 
scheme which has been adopted.
At this point the derivation of the full expression for the amplitude
$E_{R(1,1)}$ is rather immediate and gives
\be
E_{R(1,1)} (t,\theta) = { E_{R(1,1)} (t,\theta) |_{g^2_R} \over 1 + \Pi_R (t) }
= g^2_R {1 \over t [ 1 + \Pi_R (t) ]} \cot \theta \cdot (G_1)_{ij,kl} ~.
\ee
The value of $E_{R(1,1)}$ is gauge--dependent (as $M_{R (1,1)}$ was, too), 
since the gauge parameter $\alpha_R$ does appear inside $\Pi_R$ at the 
right--hand--side of Eq. (3.61). 
For the following calculations we shall fix the gauge parameter $\alpha_R$ 
to 1 ({\it Feynman} gauge), in conformity with the choice we have made in 
the previous case.
Eq. (3.61) is the full expression for $E_{R(1,1)}$, not truncated at any 
perturbative order. Yet, we only need the espression for
$E_{R(1,1)}$ up to the order $O(g_R^4)$:
\be
E_{R(1,1)} (t,\theta) |_{g_R^4} = 
g^2_R {1 \over t} [ 1 - g_R^2 F^{(2)} (t) ] \cot \theta \cdot 
(G_1)_{ij,kl} ~.
\ee
Therefore, in the Euclidean theory, the contribution coming from the 
$O(g_R^4)$ diagrams shown in Figs. 3(l) to 3(p) is given by
\be
E^{(l)} (t,\theta) + \ldots + E^{(p)} (t,\theta) =
-g_R^4 { F^{(2)} (t) \over t } \cot \theta \cdot (G_1)_{ij,kl} ~.
\ee
The contribution of order $O(g_R^4)$ coming from the two Feynman diagrams shown 
in Fig. 2(b) (the {\it ladder} term) and Fig. 2(c) (the {\it cross} 
term), i.e., the contribution $E_{(2,2)} |_{g_R^4}$, turns out to be (in 
the {\it Feynman} gauge $\alpha_R = 1$)
\ba
\lefteqn{
E^{(b)} (t,\theta) + E^{(c)} (t,\theta) = }
\nonumber \\
& & = E_{(2,2)} (t,\theta) |_{g_R^4} = 
E^{(G_1)} (t,\theta) \cdot (G_1)_{ij,kl} +
E^{(G_2)} (t,\theta) \cdot (G_2)_{ij,kl} ~,
\ea
where $G_1$ and $G_2$ are two color factors defined in Eqs. (3.15) and 
(3.24).
The coefficients $E^{(G_1)} (t,\theta)$ and $E^{(G_2)} (t,\theta)$ 
in front of the color factors in Eq. (3.64) are found to be
\ba
E^{(G_1)} (t,\theta) &=& {N_c g_R^4 \over 4 \pi} I(t)
\theta \cot^2 \theta ~; \nonumber \\
E^{(G_2)} (t,\theta) &=& {1 \over 2} g_R^4 I(t)
\cot^2 \theta ~.
\ea
The contribution $E_{(3,1)} |_{g_R^4} + E_{(1,3)} |_{g_R^4}$ from the diagrams 
in Figs. 2(d) to 2(i), obtained after expanding one of the two Wilson lines 
up to the order $O(g^3)$, and the remaining one up to the first order in $g$, 
can be written as
\ba
E_{(3,1)} |_{g_R^4} &=& 
Z_{E W}^{(2)} g_R^2 \cdot E_{R(1,1)} |_{g_R^2} + \Delta E_{(3,1)} ~,
\nonumber \\
E_{(1,3)} |_{g_R^4} &=& 
Z_{E W}^{(2)} g_R^2 \cdot E_{R(1,1)} |_{g_R^2} + \Delta E_{(1,3)} ~,
\ea
where $\Delta E_{(3,1)}$ and $\Delta E_{(1,3)}$ are divergent quantities, 
whose regularized expressions depend on the adopted renormalization scheme. 
In the MS renormalization scheme one finds that
\be
\Delta E_{(3,1)} = \Delta E_{(1,3)} = 
E_{R(1,1)} |_{g^2_R} \cdot {g^2_R \over (4\pi)^2}
N_c \left[ {1 \over \varepsilon} + B \right] ~,
\ee
where $B$ is the same finite number (as $\varepsilon$ goes to zero), which 
appears in the corresponding expression (3.29) for the Minkowskian case.
From Eq. (3.67) and from the expression (3.30) of $Z_{1 W}$ up to the order
$O(g_R^2)$, one immediately derives that
\be
\Delta E_{(3,1)} + \Delta E_{(1,3)} 
+ 2 Z_{1 W}^{(2)} g_R^2 \cdot E_{R(1,1)} |_{g_R^2} =
E_{R(1,1)} |_{g^2_R} \cdot {g^2_R \over (4\pi)^2} 2 N_c B ~.
\ee
As in the Minkowskian case, the divergence contained in $\Delta E_{(3,1)}
+ \Delta E_{(1,3)}$ is exactly 
cancelled out by the two diagrams with the counterterm $Z_{1 W}^{(2)}$,
represented in Figs. 3(q) and 3(r).
Finally, one has to evaluate the contributions $E_{(2,1)} |_{g_R^4}$ and 
$E_{(1,2)} |_{g_R^4}$, represented by the two diagrams in Figs. 3(j) and 3(k), 
respectively. Again, explicit calculations show 
that their contribution vanishes:
\be
E_{(2,1)} |_{g_R^4} = E_{(1,2)} |_{g_R^4} = 0 ~.
\ee
We can now sum up all the contributions previously evaluated in 
order to find the complete expression for the amplitude $E$, defined by
Eq. (3.50), up to the order $O(g_R^4)$:
\be
E (t,\theta) |_{g_R^4} =
\left[ 1 + \left( 2 Z_{E W}^{(2)}
- F^{(2)} (t) + {2 N_c B  \over (4\pi)^2} \right) g^2_R \right] 
\cdot E_{R(1,1)} (t,\theta) |_{g^2_R} + E_{R(2,2)} (t,\theta) ~.
\ee
Introducing here the expressions found above for $E_{R(1,1)} |_{g_R^2}$ 
[see Eq. (3.55)] and for $E_{R(2,2)} |_{g_R^4}$ [see Eqs. (3.64) and (3.65)], 
we finally find the following expression for $E (t,\theta) |_{g_R^4}$:
\ba
\lefteqn{
E (t,\theta) |_{g_R^4} = g^2_R {1 \over t} \cot \theta }
\nonumber \\
& & \times \left[ 1 + \left( 2 Z_{E W}^{(2)} - F^{(2)} (t) + 
{2 N_c B \over (4\pi)^2} + {N_c \over 4\pi} t I(t) \theta \cot \theta
\right) g^2_R \right] \cdot (G_1)_{ij,kl}
\nonumber \\
& & + {1 \over 2} g_R^4 I(t) \cot^2 \theta \cdot (G_2)_{ij,kl} ~.
\ea
The quantity $g_E (t,\theta)$, defined by Eq. (3.41), can be immediately 
derived up to the order $O(g_R^4)$, making use also of
the expansion (3.45) for the renormalization constant $Z_{E W}$:
\ba
\lefteqn{
g_E (t,\theta) |_{g_R^4} = { E (t,\theta) \over Z_{E W}^2 } 
\displaystyle\vert_{g_R^4} = 
E (t,\theta) |_{g_R^4} - 2 Z_{E W}^{(2)} g_R^2 
\cdot E (t,\theta) |_{g^2_R} }
\nonumber \\
& & = g^2_R {1 \over t} \cot \theta
\left[ 1 - \left( F^{(2)} (t) + {2 N_c B \over (4\pi)^2} + 
{N_c \over 4\pi} t I(t) \theta \cot \theta \right) g^2_R \right] 
\cdot (G_1)_{ij,kl}
\nonumber \\
& & + {1 \over 2} g_R^4 I(t) \cot^2 \theta \cdot (G_2)_{ij,kl} ~.
\ea
After comparing the two expressions (3.35) and (3.72) 
for $g_M (t, \chi) |_{g_R^4}$ and $g_E (t, \theta) |_{g_R^4}$, 
we immediately recognize that they are 
linked by the following analytic continuation in the angular variable:
\ba
g_E (t, \theta) |_{g_R^4}
\mathop{\longrightarrow}_{\theta \to -i \chi}
g_E (t, -i \chi) |_{g_R^4} = g_M (t, \chi) |_{g_R^4} ~;
\nonumber \\
{\rm or:}~~ g_M (t, \chi) |_{g_R^4}
\mathop{\longrightarrow}_{\chi \to i \theta}
g_M (t, i \theta) |_{g_R^4} = g_E (t, \theta) |_{g_R^4} ~.
\ea
This is the same relation we have already found in the preceding section for 
the corresponding quantities in the Abelian case. It appears to be an 
absolutely ``{\it natural}'' correspondence law. There is apparently no 
reason why it should not be true at higher perturbative orders: we shall 
therefore assume that it is valid for the ``{\it full}'' amplitudes, i.e.,
not truncated at any perturbative order. In the next section 
we shall discuss some interesting consequences and some possible 
applications of this relationship of analytic continuation.

\newsection{Concluding remarks and prospects}

\noindent
In the preceding section we have seen that also for a non--Abelian 
gauge theory it is possible to reconstruct the high--energy scattering 
amplitude by evaluating a correlation of infinite Wilson lines forming a 
certain angle $\theta$ in Euclidean four--space, then by continuing this 
quantity in the angular variable, $\theta \to -i \chi$, where $\chi$ is 
the hyperbolic angle between the two Wilson lines in Minkowski 
space--time, and finally by performing the limit $\chi \to \infty$ (i.e.,
$\beta \to 1$). In fact, the high--energy scattering amplitude is given by
\ba
\lefteqn{
M_{fi} = \langle \psi_{i\alpha}(p'_1) \psi_{k\gamma}(p'_2) | M | 
\psi_{j\beta}(p_1) \psi_{l\delta}(p_2) \rangle } \nonumber \\
& & \mathop{\sim}_{s \to \infty}
-i \cdot 2s \cdot \delta_{\alpha\beta} \delta_{\gamma\delta}
\cdot g_M (t,\chi \to \infty) ~.
\ea
The quantity $g_M (t,\chi)$, defined by Eq. (3.1) in the Minkowski world, is 
linked to the corresponding quantity $g_E (t,\theta)$, defined by Eq. (3.41) 
in the Euclidean world, by the following analytic continuation in the 
angular variables:
\ba
g_E (t,\theta)
\mathop{\longrightarrow}_{\theta \to -i \chi}
g_E(t,-i \chi) = g_M (t,\chi) ~;
\nonumber \\
{\rm or:}~~ g_M (t,\chi)
\mathop{\longrightarrow}_{\chi \to i \theta}
g_M (t,i\theta) = g_E (t,\theta)
~.
\ea
The important thing to note here is that the quantity $g_E (t,\theta)$, 
defined in the Euclidean world, may be computed non perturbatively 
by well--known and well--established techniques: first of all, of course,
by means of the formulation of the theory on the lattice. Also the 
stochastic vacuum model \cite{Dosch},
which is naturally defined for the Euclidean 
theory, may provide a suitable instrument to evaluate the quantity (3.41).
In all cases, once one has obtained the quantity $g_E (t,\theta)$, one 
still has to perform an analytic continuation in the angular variable
$\theta \to -i \chi$, and finally one has to extrapolate to the limit 
$\chi \to \infty$ (i.e., $\beta \to 1$). 
We are fully aware that this may not be an easy way.
Nevertheless, interesting new results are expected along this direction.
As an example, we shall show how, using this 
approach, one can re--derive the well--known {\it Regge Pole Model}
\cite{Regge}, but in a different way, with respect to the 
original derivation.
First of all, we write $g_E (t,\theta)$ in the partial--wave expansion:
\be
g_E (t,\theta) = \displaystyle\sum_{l=0}^{\infty} A_l (t) P_l (\cos \theta)
~.
\ee
If $A_l (t)$ can be analytically continued to complex values of $l$, then 
we can re--write Eq. (4.3) in the following way:
\be
g_E (t,\theta) = {1 \over 2i} \displaystyle\int_C
{ A_l (t) P_l (-\cos \theta) \over \sin (\pi l) } dl ~,
\ee
where $C$ is a contour in the complex $l$--plane, running anti--clockwise 
around the real positive l--axis and enclosing all non--negative integers, 
while excluding all the singularities of $A_l$.
Eq. (4.4) can be verified after recognizing that $P_l (-\cos \theta)$ is an 
integer function of $l$ and that the singularities enclosed by the contour 
$C$ of the expression under integration in the Eq. (4.4) are simple poles at 
the non--negative integer values of $l$.
So the right--hand side of (4.4) is equal to the sum of the residues of the 
integrand in these poles and this gives exactly the right--hand side of 
(4.3). The ``{\it minus}'' sign in the argument of the Legendre function $P_l$ 
into Eq. (4.4) is due to the following relation, valid for integer values of
$l$:
\be
P_l (-\cos \theta) = (-1)^l P_l (\cos \theta) ~.
\ee
Then, we can reshape the contour $C$ into the straight line $\Re (l)
= -{1 \over 2}$. Eq. (4.4) then becomes
\be
g_E (t,\theta) = - \displaystyle\sum_{ \Re (\alpha_n) >  -{1 \over 2} }
{ \pi r_n (t) P_{\alpha_n (t)} (-\cos \theta) \over \sin (\pi \alpha_n (t))}
- {1 \over 2i} \displaystyle\int_{-{1 \over 2} -i\infty}^{-{1 \over 2} 
+i\infty} { A_l (t) P_l (-\cos \theta) \over \sin (\pi l) } dl ~,
\ee
where $\alpha_n (t)$ is a pole of $A_l (t)$ in the complex $l$--plane and
$r_n (t)$ is the corresponding residue. We have assumed that $A_l$ vanishes
enough rapidly as $|l| \to \infty$ in the right half--plane, so that the 
contribution from the infinite contour is zero.
Eq. (4.6) immediately leads to the asymptotic behavior of the scattering 
amplitude in the limit $s \to \infty$, with a fixed $t$ ($|t| \ll s$).
In fact, making use of the analytic extension (4.2) when continuing the 
angular variable, $\theta \to -i\chi$, we derive that
\ba
\lefteqn{
g_M (t,\chi) = g_E (t,-i\chi) 
} \nonumber \\
& & = - \displaystyle\sum_{ \Re (\alpha_n) >  -{1 \over 2} }
{ \pi r_n (t) P_{\alpha_n (t)} (-\cosh \chi) \over \sin (\pi \alpha_n (t))}
- {1 \over 2i} \displaystyle\int_{-{1 \over 2} -i\infty}^{-{1 \over 2} 
+i\infty} { A_l (t) P_l (-\cosh \chi) \over \sin (\pi l) } dl ~.
\ea
The hyperbolic angle $\chi$ is linked to $s$ by the relation (3.39).
Therefore we can re--express $\cosh \chi$ in terms of $s$ in the following 
way:
\be
\cosh \chi = {s \over 2m^2} - 1 ~.
\ee
The asymptotic form of $P_\alpha (z)$ when $|z| \to \infty$ is well known.
It is a linear combination of $z^\alpha$ and of $z^{-\alpha -1}$.
When $\Re (\alpha) >  -1/2$, this last term can be neglected.
Therefore, in the limit $s \to \infty$, with a fixed $t$ ($|t| \ll s$),
we are left with the following expression:
\be
g_M (t,\chi \to \infty)
\sim \displaystyle\sum_{ \Re (\alpha_n) >  -{1 \over 2} }
{ \beta_n (t) s^{\alpha_n (t)} \over \sin (\pi \alpha_n (t))} ~,
\ee
where $\beta_n (t)$ is independent on $s$ (it only depends on $t$).
The integral in Eq. (4.7), usually called the {\it background term}, vanishes 
at least as $s^{-1/2}$. Eq. (4.9) allows to immediately extract the scattering 
amplitude according to Eq. (4.1):
\ba
\lefteqn{
M_{fi} \mathop{\sim}_{s \to \infty}
-i \cdot 2s \cdot \delta_{\alpha\beta} \delta_{\gamma\delta}
\cdot g_M (t,\chi \to \infty) } \nonumber \\
& & \sim \delta_{\alpha\beta} \delta_{\gamma\delta}
\displaystyle\sum_{ \Re (\alpha_n) > -{1 \over 2} }
{ \tilde{\beta}_n (t) s^{1+\alpha_n (t)} \over \sin (\pi \alpha_n (t))} ~.
\ea
This equation gives the explicit dependence of the scattering amplitude at 
very high energy $s \to \infty$ and a fixed transferred momentum $t$
($|t| \ll s$). As we can see, this amplitude comes out to be a sum of powers 
of $s$. If we put $\bar{\alpha} (t) = 1 + \alpha_{\bar{n}} (t)$, where 
$\alpha_{\bar{n}} (t)$ is the pole with the largest real part (at that 
given $t$), we can also write
\be
M_{fi} \sim \delta_{\alpha\beta} \delta_{\gamma\delta} \cdot
\bar{\beta} (t) s^{\bar{\alpha} (t)} ~.
\ee
This sort of behavior for the scattering amplitude was first proposed by 
Regge in \cite{Regge} and $\bar{\alpha} (t)$ is often called
a ``{\it Regge pole}''. In the original derivation \cite{Regge}, 
the asymptotic behavior (4.11) was recovered by analytically 
continuing to very large imaginary values the angle between the
trajectories of the two exiting particles in the $t$--channel process.
Instead, in our derivation, we have analytically continued the quantity 
(3.41), 
defined in the Euclidean theory, to very large (negative) imaginary values 
of the angle $\theta$ between the two Euclidean Wilson lines.
As in the original derivation, we have assumed that the singularities of
$A_l$ are simple poles. If there are other kinds of singularities, 
different from simple poles, their contribution will be of a different type
and, in general, also logarithmic terms (of $s$) may appear in the
amplitude. Only a precise evaluation of $g_E (t,\theta)$ can reveal such 
behaviors (after the analytic continuation). 
In the preceding section this was done up to 
the order $O(g_R^4)$ in perturbation theory. New interesting results are 
expected from a non perturbative approach, for example by directly computing 
$g_E (t,\theta)$ on the lattice or by means of the stochastic vacuum model.

\bigskip
\noindent {\bf Acknowledgements}
\smallskip

I would like to thank Adriano Di Giacomo and G{\"u}nther Dosch 
for many useful discussions.

\vfill\eject

{\renewcommand{\Large}{\normalsize}
}

\vfill\eject

\noindent
\begin{center}
{\bf FIGURE CAPTIONS}
\end{center}
\vskip 0.5 cm
\begin{itemize}
\item [\bf Fig.~2.] The contributions 
of the type $(1,1) |_{g_R^2}$ [(a)], $(2,2) |_{g_R^4}$ [(b) and (c)], 
$(3,1) |_{g_R^4}$ [from (d) to (f)] and $(1,3) |_{g_R^4}$ [from (g) to (i)]
to the amplitudes (3.8) and (3.50). The notation is explained in the text.
\bigskip
\item [\bf Fig.~3.] The contributions of the
type $(2,1) |_{g_R^4}$ [(j)], $(1,2) |_{g_R^4}$ [(k)], 
$(1,1)$ of order $O(g_R^4)$ [from (l) to (p)], plus the counterterms 
[(q) and (r)], for the amplitudes (3.8) and (3.50). 
The notation is explained in the text.
\end{itemize}

\vfill\eject

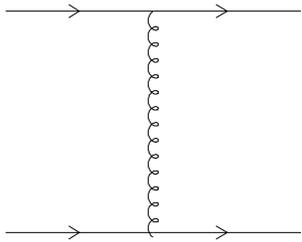
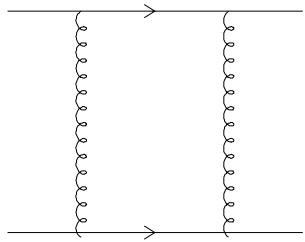
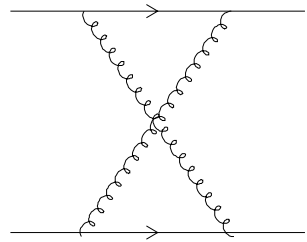
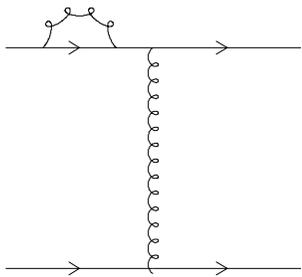
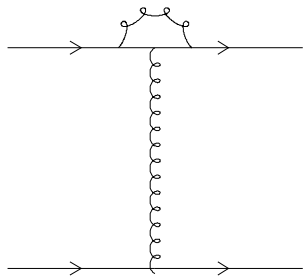
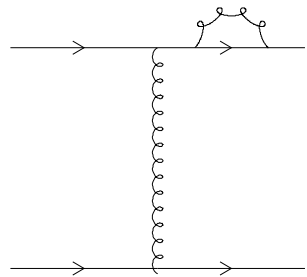
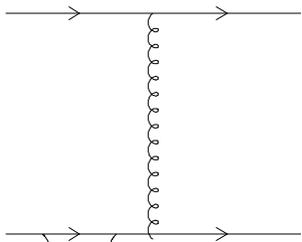
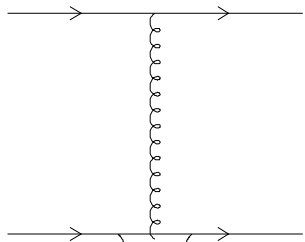
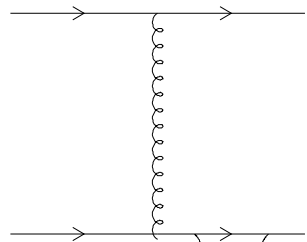
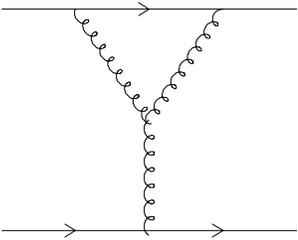
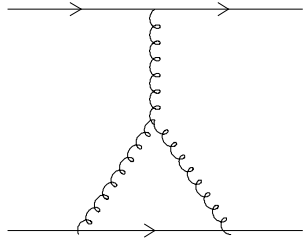
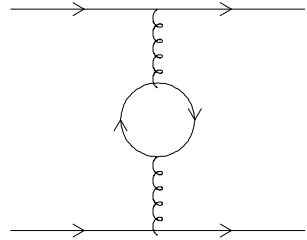
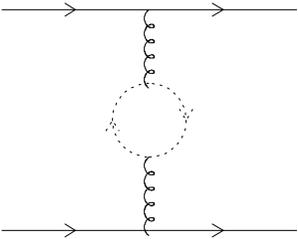
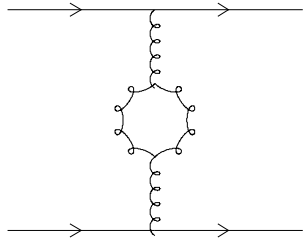
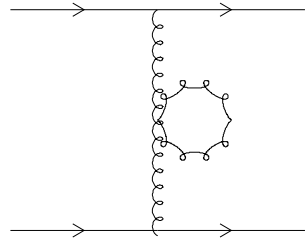
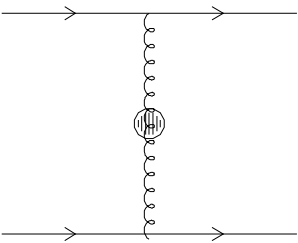
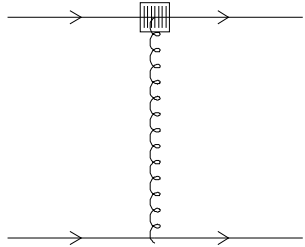
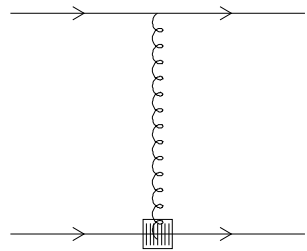
\end{document}